\documentclass[useAMS,usenatbib]{mn2e}
\usepackage{aas_macros,graphicx,times,multirow,amsmath,amssymb,color,longtable}
\title[An 1.2-s X-ray pulsar in M\,31]{EXTraS discovery of an 1.2-s X-ray pulsar in M\,31}
\author[P.~Esposito et al.] {P.~Esposito,$^{1}$\thanks{E-mail: paoloesp@iasf-milano.inaf.it} G.~L.~Israel,$^{2}$ A.~Belfiore,$^{1}$ G.~Novara,$^{3}$ L.~Sidoli,$^{1}$ G.~A.~Rodr\'iguez Castillo,$^{2}$
\newauthor A.~De~Luca,$^{1}$ A.~Tiengo,$^{1,3,4}$ F.~Haberl,$^{5}$ R.~Salvaterra,$^{1}$ A.~M.~Read,$^6$ D.~Salvetti,$^1$ S.~Sandrelli,$^7$
\newauthor  M.~Marelli,$^{1}$ J.~Wilms$^8$ and D.~D'Agostino$^9$
\smallskip\\
$^1$INAF--Istituto di Astrofisica Spaziale e Fisica Cosmica - Milano, via E. Bassini 15, I-20133 Milano, Italy\\
$^2$INAF--Osservatorio Astronomico di Roma, via Frascati 33, I-00040 Monteporzio Catone, Italy\\
$^3$IUSS--Istituto Universitario di Studi Superiori, piazza della Vittoria 15, I-27100 Pavia, Italy\\
$^4$INFN--Istituto Nazionale di Fisica Nucleare, Sezione di Pavia, via A. Bassi 6, I-27100 Pavia, Italy\\
$^5$Max-Planck-Institut f\"ur extraterrestrische Physik, Giessenbachstra{\ss}e, D-85748 Garching, Germany\\
$^6$Department of Physics and Astronomy, University of Leicester, Leicester LE1 7RH, U.K.\\
$^7$INAF--Osservatorio Astronomico di Brera, via Brera 28, I-20121 Milano, Italy\\
$^8$ECAP--Erlangen Centre for Astroparticle Physics, Sternwartstrasse 7, D-96049 Bamberg, Germany\\
$^9$CNR--Istituto di Matematica Applicata e Tecnologie Informatiche, via de Marini 6, I-16149 Genova, Italy
}
\date{Accepted 2015 December 1.  Received 2015 November 25; in original form 2015 November 13} \pagerange{\pageref{firstpage}--\pageref{lastpage}} \pubyear{2015}

\def\LaTeX{L\kern-.36em\raise.3ex\hbox{a}\kern-.15em
    T\kern-.1667em\lower.7ex\hbox{E}\kern-.125emX}
\def\xmm {\emph{XMM--Newton}}
\def\cxo {\emph{Chandra}}

\def\hst {\emph{HST}}
\def\src {\mbox{3X\,J0043}}
\def\flux {\mbox{erg cm$^{-2}$ s$^{-1}$}}
\def\lum {\mbox{erg s$^{-1}$}}
\def\nh {$N_{\rm H}$}

\begin{document}

\label{firstpage}
\maketitle
\begin{abstract}
During a search for coherent signals in the X-ray archival data of \xmm, we discovered a modulation at 1.2~s in 3XMM\,J004301.4+413017 (\src), a source lying in the direction of an external arm of M\,31. This short period indicates a neutron star (NS). Between 2000 and 2013, the position of \src\ was imaged by public \xmm\ observations 35 times. The analysis of these data allowed us to detect an orbital modulation at 1.27~d and study the long-term properties of the source. The emission of the pulsar was rather hard (most spectra are described by a power law with $\Gamma<1$) and, assuming the distance to M\,31, the 0.3--10~keV luminosity was variable, from $\sim$$3\times10^{37}$ to $2\times10^{38}$~\lum. The analysis of optical data shows that, while \src\ is likely associated to a globular cluster in M\,31, a counterpart with $V\gtrsim22$ outside the cluster cannot be excluded. Considering our findings, there are two main viable scenarios for \src: a peculiar low-mass X-ray binary, similar to 4U\,1822--37 or 4U\,1626--67, or an intermediate-mass X-ray binary resembling Her\,X-1. Regardless of the exact nature of the system, \src\ is the first accreting NS in M\,31 in which the spin period has been detected.
\end{abstract}


\section{Introduction}
EXTraS (Exploring the X-ray Transient and variable Sky) is a project to explore systematically the serendipitous content of the \xmm\ European Photon Imaging Camera (EPIC) pn \citep{struder01} and MOS \citep{turner01} data in the temporal domain. The results  will be released to the astronomical community in an easy-to-use form. The project includes a search for fast transients missed by standard image analysis, as well as the search and characterisation of variability (both periodical and aperiodical) in hundreds of thousands of sources, spanning more than nine orders of magnitude in time scale (from $<$1~s to $>$10~yr) and six orders of magnitude in flux (from $10^{-9}$ to $10^{-15}$~\flux in 0.2--12 keV). See \citet{deluca15} or the project web site, www.extras-fp7.eu, for details. \\
\indent At the moment of writing, about 7,400 \xmm\ observations were retrieved and around 1.1~million time series from sources detected with the EPIC CCDs  in imaging mode were searched for periodic signals in a systematic and automatised way with the detection algorithm described in \citet{israel96}.
Among dozens of new X-ray pulsators found so far with periodic signals at high confidence ($>$4.5$\sigma$), there is 3XMM\,J004301.4+413017 (\src) in M\,31. \src\ \citep{supper01,kaaret02} lies in the direction of an external arm to the north-east of the galaxy, and was proposed as a high-mass X-ray binary (HMXB) by \citet{shaw09}. The suggestion was based on its hard X-ray spectrum and the coincidence (within 0.7~arcsec) with a $V = 17.2$ object found in the optical catalogue by \citet{massey06}. However, based on the proximity of the source to the M\,31 globular cluster GlC\,377 \citep*{kaaret02,trudolyubov04,pietsch05}, \citet{stiele11} observed a low-mass X-ray binary (LMXB) in hard state would be more likely.\\
\indent Here we report on the discovery in \src\ of a period of 1.2~s and an orbital modulation at 1.27~d. These findings clearly indicate a neutron star (NS) in a binary system. This is the first NS in M\,31 for which a spin period has been detected. Using all the public \xmm\ observations (Sect.\,\ref{observations}), as well as optical data (Sect.\,\ref{optical}), we discuss the nature of this new X-ray pulsar (Sect.\,\ref{discussion}).

\section{\xmm\ observations and analysis}\label{observations}

The region of \src\ was repeatedly observed with \xmm\ with the EPIC detectors in full imaging mode (Full Frame). The source was always off-axis, at an angle varying from $\sim$2 to 16~arcmin, but by about 15~arcmin in most observations. Since the time resolution of the MOS cameras (2.6~s) is not adequate to sample the 1.2-s pulsation and in most pointings the source fell out of the field of view of the MOSs, we used only the pn data (readout time: 73~ms). The public pn data sets covering the position of \src\ (apart from a few in which the source was located in a CCD gap and no useful data were collected) are summarised in Table\,\ref{obslog}. They span from December 2000 to February 2013.
\begin{table}
\centering \caption{Logbook of the \xmm\ observations used in this work. The full table is available online (see Supporting Information).} \label{obslog}
\begin{tabular}{@{}lccccc}
\hline
Obs.\,ID & Start date & \multicolumn{2}{c}{Exposure$^{a}$}  & Off-axis$^{b}$ & Count rate$^{c}$ \\
&  & \multicolumn{2}{c}{(ks)} & (arcmin) & ($10^{-2}$~s$^{-1}$) \\
\hline
0112570601 & 28-12-2000 & 13.3 & (8.8)  & 15.8 &$1.97\pm0.17$   \\
0112570101 & 06-01-2002 & 64.3 & (48.3) & 15.9 &$2.58\pm0.08$   \\
0402560901 & 26-12-2006 & 61.9 & (38.3) & 14.7 & $3.39\pm0.10$  \\
0405320701 & 31-12-2006 & 15.9 & (12.2) & 15.5 & $3.47\pm0.19$  \\
0405320801 & 16-01-2007 & 13.9 & (10.5) & 15.5 & $3.76\pm0.21$  \\
\hline
\end{tabular}
\begin{list}{}{}
\item[$^{a}$] In parentheses we give the good observing time after dead-time correction and screening for soft proton flares.
\item[$^{b}$] Radial off-axis angle of \src\ from the boresight of the pn telescope.
\item[$^{c}$] Net source count rate in the 0.3--10~keV energy band using the extraction regions described in the text; the values are not corrected for point spread function and vignetting effects.
\end{list}
\end{table}

The raw observation data files (ODF) retrieved from the \xmm\ Science Archive were processed with the Science Analysis Software (\textsc{sas}) v.14. The screening of time periods with high particle background was based on the good-time intervals included in the processed pipeline products (PPS). 
We extracted the event lists and spectra using an extraction radius of 20~arcsec, and estimated the background from regions near the source, with radius of 25~arcsec and avoiding CCD gaps. 
To convert the event times to the barycentre of the Solar System, we processed the event files with the \textsc{sas} task \textsc{barycen} using the source position. Spectra were rebinned so as to obtain a minimum of 30 counts per energy bin, and for each spectrum we generated the response matrix and the ancillary file using the \textsc{sas} tasks \textsc{rmfgen} and \textsc{arfgen}.

\subsection{Discovery of the period and timing analysis}\label{timing}

\begin{figure}
\centering
\resizebox{\hsize}{!}{\includegraphics[angle=-90]{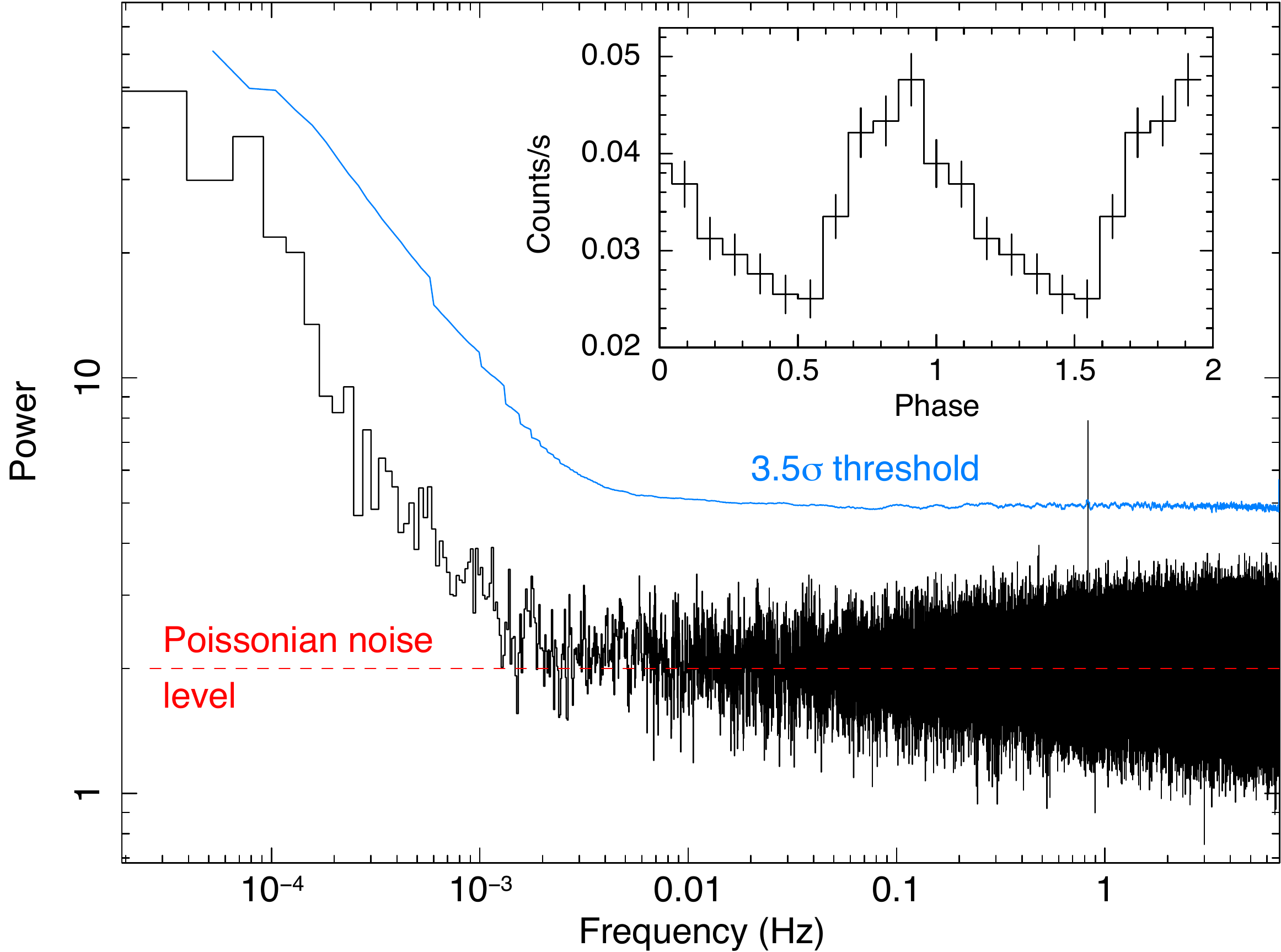}}
\caption{\label{fft_efold} Fourier power spectrum from the December 2000--February 2013 pn data (0.5--10~keV; average of 36 Fourier transforms; $\Delta t\simeq73$~ms, 262,144 frequencies). The blue line corresponds to the 3.5$\sigma$ confidence level threshold for potential signals and was computed taking into account the number of trials, equal to the number of frequency bins of the spectrum. The prominent peak above the threshold corresponds to the 1.2-s signal. The inset shows the light curve of the data set 0650560301 (where the signal was discovered) folded to its best period.}
\end{figure}

A periodic signal at about 1.2~s was first detected at a confidence level of about 6.5$\sigma$ by the automatic analysis in the pn data of obs. 0650560301 (see Table\,\ref{obslog}). By a $Z^2_1$ (or Rayleigh test), the value was refined to $1.203830\pm0.000003$~s; the corresponding pulse profile is single-peaked, with a substantial pulsed fraction [($50\pm2\%$), root mean square (RMS)]. Subsequently, we computed a power spectrum using the whole available data set (for a timespan of about 12 years) and this confirmed the presence of the 1.2-s signal at a confidence greater than 12$\sigma$ (Fig.\,\ref{fft_efold}). 

In each observation, we measured the period by the $Z^2_1$ test. A constant fit shows that the periods are not consistent with a single value, with a reduced $\chi^2$ ($\chi^2_\nu$) higher than 200 for 33 degrees of freedom (dof). The inclusion of a period derivative component (or of higher-order derivatives) does not give better fits. On the other hand, the inspection of the longest observations reveals a strong modulation of the pulsar period, which is consistent with Doppler shifts induced by an orbital motion with period of $\approx$1~d.

To model the binary parameters, we began by dividing the data into $\sim$30-ks-long segments. We searched each segment for coherent pulsations around the spin period, through a polynomial fit with \textsc{presto} \citep{ransom01}. We could collect good quality time of arrival of the pulses (TOAs) with a cadence of 1.5--3~ks. Obs. 0690600401 (see Table\,\ref{obslog}) covers a full orbit and we could infer a preliminary binary solution, by fitting the orbital parameters together with the spin period with \textsc{tempo2} \citep{hobbs06} using the binary model \textsc{ell1} \citep{lange01}. We extended the timing solution to obs. 0112570101, taken more than 10 years before, slightly refining the orbital period estimate, and fitting for the spin period at the epoch of the observation. Because this estimate of the orbital period suffers from aliasing, we broke this degeneracy by analysing in the same way also obs. 0650560301. The orbital solution, reported in Table\,\ref{bpar}, produces residuals of 80--100~ms (RMS) in each observation. The spin periods measured at the 3 epochs are $1.203892\pm0.000001$~s on MJD 52281, $1.203644\pm0.000003$~s on MJD 55566, and $1.2037007\pm0.0000003$~s on MJD 56105, indicating alternated trends of spin-up and \mbox{-down}. A detailed timing analysis of \src\ will be addressed in a dedicated forthcoming publication. We finally notice that no significant flux variations are observed along the orbit, with a 3$\sigma$ upper limit on the modulation amplitude set from obs. 0690600401 at 10\%. \begin{table}
\centering \caption{Orbital parameters of \src.} \label{bpar}
\begin{tabular}{@{}lr}
\hline
Parameter & Value\\
\hline
Orbital period, $P_{\rm b}$ (d) & $1.27397828\pm0.00000071$\\
Epoch of ascending node, $T_{\rm asc}$ (MJD) & $56104.7912\pm0.0011$ \\
Projected semi-axis, $A_{\rm X}\sin i$ (lt-s) & $2.884\pm0.017$ \\
Eccentricity, $e$ & $0.011\pm0.009$$^a$\\
Longitude of periastron, $\omega$ (\degr) & $276\pm41$ \\
Mass function (M$_{\sun}$) & $0.0159\pm0.0008$\\
Minimum companion mass$^b$ (M$_{\sun}$) & 0.36 \\
\hline
\end{tabular}
\begin{list}{}{}
\item[$^{a}$] Upper limit at the 3$\sigma$ confidence level: $e<0.037$.
\item[$^{b}$] Value computed for an orbit viewed edge-on, $i=90\degr$.
\end{list}
\end{table}

\subsection{Spectral analysis and long-term variability}\label{spectroscopy}
The spectral fitting was performed in 0.3--10~keV using \textsc{xspec} v.12.8; the abundances used are those of \citet*{wilms00}. Using \xmm\ data, \citet{trudolyubov04} observed that the spectrum of \src\ could be described by a hard power law modified for the interstellar absorption with an equivalent hydrogen column depth $N_{\rm H}=7\times10^{20}$~cm$^{-2}$, which is the typical value in the direction of M\,31.\\
\indent The power-law model indeed works for most spectra, but when applied to data sets 0690600401 and 0700380601, it overpredicts the emission above 7--8~keV, resulting in poor fits ($\chi^2_\nu=1.30$ for 157 dof and 1.34 for 32 dof, respectively). It should be noticed that these are the only observations with both small off-axis angle and good count-statistics (Table\,\ref{obslog}).
In particular, the spectrum from obs. 0690600401, which is the richest one, with about 4,500 net counts, is better described by a power-law with an exponential cutoff, or with the addition of a thermal component. Adopting a cutoff power law ($\chi^2_\nu=1.05$ for 155 dof), the photon index is $\Gamma=0.0\pm0.1$ and the cutoff is located at $3.6^{+0.4}_{-0.3}$~keV; the absorption is poorly constrained, with a 3$\sigma$ upper limit $N_{\rm H}<7\times10^{20}$~cm$^{-2}$, and the observed flux is $(4.8\pm0.2)\times10^{-13}$~\flux. For the distance to M\,31 ($d= 780$~kpc; \citealt{holland98}), this translates into a luminosity of $(3.6\pm0.1)\times10^{37}$~\lum. A power law plus blackbody model yields a $\chi^2_\nu=1.09$ for 154 dof, and the parameters are: $N_{\rm H}=(7\pm2)\times10^{20}$~cm$^{-2}$, $\Gamma=1.0^{+0.2}_{-0.1}$, $kT=1.3^{+0.2}_{-0.1}$~keV and a blackbody radius $R_{\rm BB} = 6.5^{+0.9}_{-0.7}$~km (at 780~kpc); the observed flux is $(4.8\pm0.2)\times10^{-13}$~\flux\ ($\sim$40\% of which from the blackbody component) and the luminosity is $(3.7\pm0.1)\times10^{37}$~\lum. Similar results are obtained for obs. 0700380601.\\
\indent The automatic analysis of long-term variability performed within the EXTraS project (assuming a constant spectral shape) shows significant changes in flux ($\chi^2_\nu>80$ for 34 dof for a constant fit), with a factor $>$2 drop in flux in approximately a year, starting around January/February 2011. We assessed by a `runs' (Wald--Wolfowitz) test that these changes are not consistent with an unchanging underlying distribution.
\begin{figure*}
\centering
\resizebox{.97\hsize}{!}{\includegraphics[angle=0]{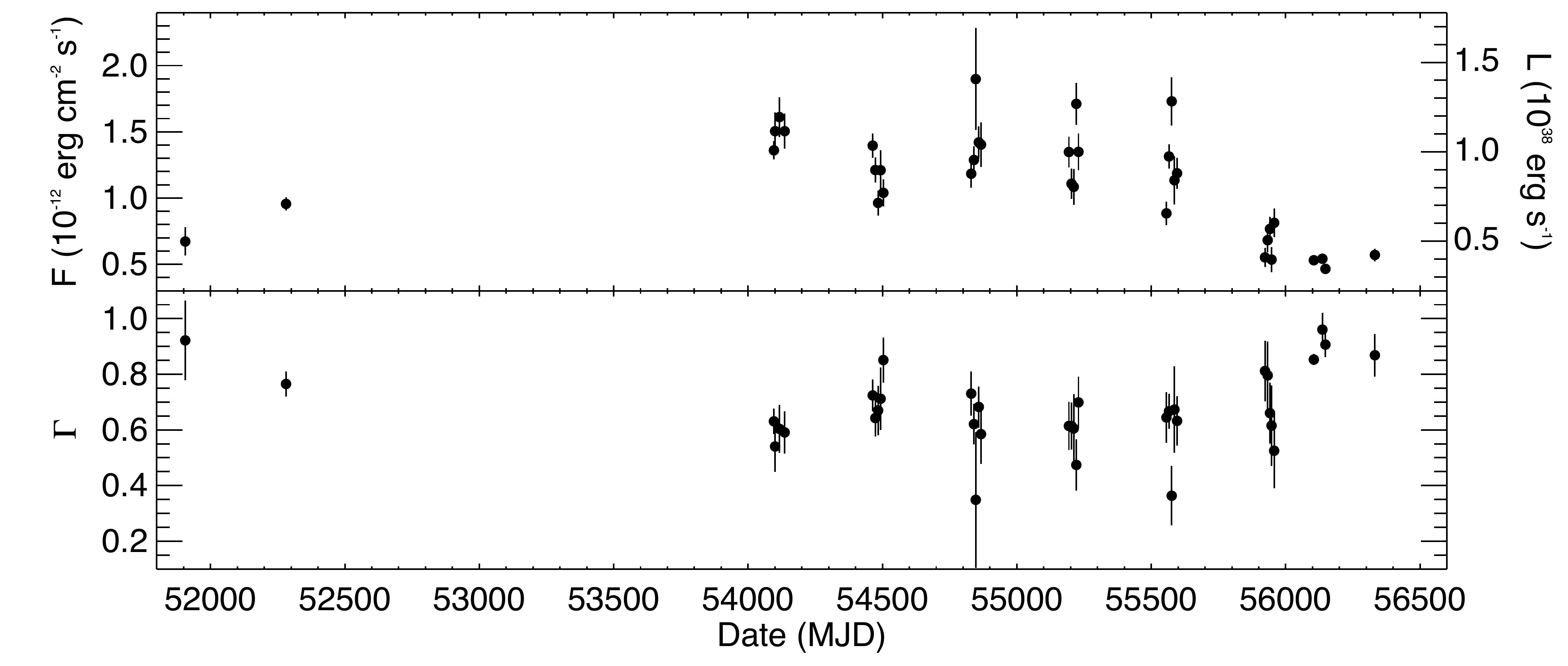}}
\caption{\label{lightcurve} Long-term properties of \src\ as observed with \xmm. The flux (upper panel) and the photon index (lower panel) were derived in the 0.3--10~keV range from the power-law fit of all data sets with the \nh\ fixed at the value towards M\,31 using the \textsc{cflux} model in \textsc{xspec}. In the upper panel, the right axis indicates the isotropic 0.3--10~keV luminosity assuming a distance of 780~kpc; to convert from observed flux to (unabsorbed) luminosity, we used a single conversion factor, which is precise to within 2\% for each observation.}
\end{figure*}
\\
\indent We further investigated long-term variability by performing a more detailed spectral analysis. \src\ displays significant spectral variations
 over the 2000--2013 period. In a simultaneous fit of all spectra with an absorbed power-law model, the variability cannot be accounted for by leaving free to vary in each data set only one of the parameters: either the absorption, the photon index, or the normalisation. Data are well fit by a power-law model with the \nh\ fixed at the M\,31 value, when photon indices and normalisations are free to vary ($\chi^2_\nu=1.16$ for 670 dof; excluding obs. 0690600401 and 0700380601, the fit is even better, with $\chi^2_\nu=1.03$ for 478 dof). The long-term light curve from the fluxes derived with this simple model is plotted in Fig.\,\ref{lightcurve}. The minimum observed flux, on 8 August 2012, was $(4.7\pm0.3)\times10^{-13}$~\flux\ in the 0.3--10~keV energy band (luminosity:  $\sim$$3.5\times10^{37}$~\lum), while the maximum was $(1.9\pm0.4)\times10^{-12}$~\flux\ (luminosity: $1.5\times10^{38}$~\lum) on 15 January 2009. The analysis also confirms the hard spectrum, with photon indices in the range from $\sim$0.35 to 1 (average value: $\Gamma=0.7$) and shows a hint of a `harder-when-brighter' correlation between flux and spectral shape. When the absorption is left free to vary but held to a common value for all data sets, the best-fitting value is $(7.5\pm0.9)\times10^{20}$~cm$^{-2}$ ($\chi^2_\nu=1.16$ for 669 dof). 

\vspace{-0.3cm}

\section{The optical field of \src}\label{optical}
The position of \src\ from the third \xmm\ Serendipitous Source Catalogue (3XMM; \citealt{rosen15}) is $\rm RA_{J2000} = 00^h43^m01\fs48,~Decl._{J2000}=+41\degr30'16\farcs9$ (error radius: about 1.5~arcsec; consistent coordinates and similar uncertainties are given by the \cxo\ Source Catalog; \citealt{evans10}). We inspected a Digitized Sky Survey image of the field of \src. A very bright source, \mbox{TYC\,2805-2136-1} ($V\sim11$), is at $\sim$9.5~arcsec (using the Tycho-2 catalogue position; \citealt{hog00}) from \src. On 20 September 2015, we obtained three spectra of this star using the 1.52-m Cassini telescope at the Loiano Observatory (Italy) equipped with the Bologna Faint Object Spectrograph \& Camera (BFOSC).
The spectra show no emission lines and allowed us to classify the object as a K0V--K1V spectral-type star. Its large distance from \src\ (also taking into account the  proper motion) rules out an association between the two objects. 

The position of \src\ was imaged on 19 February 2005 (proposal ID: 10273) by the \emph{Hubble Space Telescope} (\hst) with the Wide Field Channel (WFC) of the Advanced Camera for Surveys (ACS) instrument using the F555W filter (exposure time of 151~s; dataset J92GA7VLQ), as well as the F814W filter (exposure time of 457~s; dataset J92GA7VMQ). We retrieved geometrically corrected images produced by the on-the-fly reprocessing pipeline from the Mikulski Archive for Space Telescopes (MAST), and we performed a source detection using the \textsc{SExtractor} package \citep{bertin96}. For the detection of a source, we required a minimum of five contiguous pixels with a  threshold of 5$\sigma$ above the RMS background, with a total of 32 deblending subthresholds, and with a contrast parameter of 0.0001, setting the background mesh size to $8\times8$ ACS pixels. Using such approach, the globular cluster GlC\,377 (0.6~arcsec from \src) is undoubtedly identified amid a rather crowded field. No useful photometry could be extracted within $\sim$0.9~arcsec of the globular cluster core, due to source confusion. Outside that region, many sources consistent with the X-ray position of \src\ are detected, the brightest of which have a magnitude of $21.6\pm0.1$ in the F814W filter and of $22.1\pm0.1$ in the F555W filter (ST magnitude system).

We did not identify any star-like source possibly corresponding to the object in the catalogue by \citet{massey06}, J004301.51+413017.5, with $V=17.242$, $B$--$V= -0.655$, \mbox{$U$--$B= 1.302$}, and $V$--$R=0.370$. In the eventuality of an optical transient, we inspected the Kitt Peak National Observatory images used by  \citet{massey06}.
We conclude that the source J004301.51+413017.5 was actually a blend of the globular cluster and other sources not resolved in the data of that survey.

\vspace{-0.3cm}

\section{Discussion}\label{discussion}
We discovered in \src\ pulsations at 1.2~s, pointing to a NS, and an orbital modulation with period 1.27~d. The X-ray emission was described well enough by a hard power-law with $\Gamma< 1$. On time-scales of years, the X-ray flux varied by a factor $\approx$5, in the range $\sim$0.5--$2\times10^{-12}$~\flux. 
Eclipses or dips were not detected in the light curves, and no modulation of the flux was seen along the orbit, down to a limit of $\sim$10\%. This suggests an inclination angle $i\lesssim60\degr$.\\
\indent A chance alignment of \src\ with M\,31 cannot be ruled out but seems extremely unlikely.  The absence of any bright optical counterpart in the deep \hst\ images and the high Galactic latitude, $b=-21\degr$, far off the plane, strongly disfavour a Galactic interpretation. Furthermore, the value of the absorbing column of \src\ is very similar to the total Galactic depth in its direction. Several pieces of information also argue against a non-accreting NS, chiefly:  the periods of spin-up and \mbox{-down} and the substantial timing noise, the variability of the X-ray emission (or the unusual spectrum in the case of a magnetar), and the high luminosity of $\sim$0.4-- $2\times10^{38}$~\lum\ (at the distance of M\,31 and assuming isotropy). Therefore, we will discuss the nature of \src\ in the hypothesis of an accretion-powered source belonging to M\,31.\\
\indent \src\ lies very close to the globular cluster GlC\,377 but, considering the crowded field and the positional uncertainties, we think that the association should not be taken for granted. Apart from the unresolved core of the globular cluster, the brightest source detected in the X-ray error circle has a $V$ magnitude of 22.1. We take this value as the limit on the magnitude of the counterpart to \src. The reddening toward M\,31 ($A_V\simeq0.2$) and its distance module (24.45) implies an absolute visual magnitude, $M_V$, of the optical counterpart fainter than $-2.5$. The lower limit on the mass of the companion from the orbital parameters is $\sim$0.4~M$_{\sun}$.\\
\indent If we consider the Corbet diagram of spin versus orbital periods of accreting XRBs \citep{enoto14}, there are essentially three possibilities among known objects for an XRB with a spin period of $\sim$1.2~s in a $\sim$1~d orbital period: (i) a HMXB with a Roche-lobe-filling high-mass star;
(ii) a peculiar LMXB (similar to 4U\,1626--67 and 4U\,1822--37);
(iii) an `intermediate-mass' XRB (like \mbox{Her\,X--1}).\\ 
\indent A HMXB (i) is highly unlikely, due to the constraints from the optical photometry and the orbital parameters (a companion $>$8~M$_{\sun}$ would require an extremely low inclination, $i<9\degr$).
In the second case, (ii), \src\ could be a LMXB similar to 4U\,1822--37 (spin period: $P_{\rm s}=0.59$~s and orbital period $P_{\rm b}=0.23$~d; \citealt{jonker01}) or to the ultracompact binary 4U\,1626--67 ($P_{\rm s}=7.7$~s and $P_{\rm b}=0.03$~d; \citealt{rappaport77,chakrabarty98}).
These systems display a peculiar NS rotational period, that is much longer than expected (i.e., at milli-second level) in LMXBs in the standard picture where LMXBs are the progenitors of millisecond pulsars. They also emit X-rays with a harder (compared with the bulk of LMXBs) power-law spectrum below 10~keV,  together with a blackbody component. Their X-ray luminosity, reaching a few times $10^{37}$~\lum\ at most \citep{orlandini98,sasano14,iaria15}, is fainter than that observed in \src. However, we note that  4U\,1822--37 is an accretion disc corona source, where the true  X-ray luminosity is thought to be higher than observed, and may be at the Eddington limit for a NS (e.g. \citealt{bayless10,burderi10,iaria15}). The NS rotation period in 4U\,1626--67 shows prolonged epochs of steady spin-up and spin-down, which is also the case of \src. The pulsed fraction displayed by 4U\,1626--67 is much higher ($\sim$26--50\%, \citealt{beri14}) than the one observed in 4U\,1822--37 ($\sim$0.25\% in the energy range 2--5.4~keV, while reaching 3\% above 20~keV, \citealt{jonker01}), and more similar to that of \src. However, both sources, and 4U\,1626--67 in particular, have a much shorter orbital period. An accretion disk in a LMXB is allowed by the optical photometry, since the mean absolute magnitude expected from a disc dominating the optical emission in a LMXB can span a large range peaking at $M_V=1$ \citep{vanparadijs94}. The possible optical association with a globular cluster in M\,31 obviously favours a LMXB nature, with properties similar to those displayed by 4U\,1822--37 or 4U\,1626--67 in our Galaxy. Given the crowded field of \src, also a LMXB not associated with the globular cluster is possible.

The third possibility is that \src\ is an object similar to \mbox{Her\,X--1}, an `intermediate mass' X-ray binary (with a main-sequence A star companion) seen almost edge on, which has a spin period of 1.2~s for the NS and orbital period of 1.7~d \citep{tananbaum72}. For \src, a main sequence B and later spectral type star is a viable possibility (e.g. \citealt{schoenberner95}).  If the donor is a $\sim$2-M$_{\sun}$ star like in \mbox{Her\,X--1}, the association with a globular cluster can be ruled-out. \mbox{Her\,X--1} displays an intensity and spectral variability with a on/off cycle of 35~days. The X-ray emission from \mbox{Her\,X--1} is well described by a power-law with similar photon index as observed in \src\ (below 10~keV);
a soft excess, which can be described by a blackbody with temperature $kT\sim130$--140~eV, is also present, and is likely associated with the accretion disk (e.g. \citealt{burderi00,furst13}). Although \mbox{Her\,X--1} shows a lower X-ray luminosity than \src, reaching only $\sim$$10^{37}$~\lum, its spectral and timing properties (hard power law below 10 keV, e.g. \citealt{oosterbroek97}, and spin and orbital periods) are similar to \src, except for the 35-cycle which, at any rate, is due to the viewing angle of the system. Another argument which favors a system similar to Her\,X--1 is the very high observed luminosity, which can be produced more easily in an intermediate-mass system rather than in a LMXB, where the mass accretion rate predicted from the secular evolution is in general small \citep{verbunt93}.

We finally notice that if the NS in \src\ is spinning at, or close to, the equilibrium period (i.e. when the magnetospheric radius equals the corotation one), as it is suggested by the alternating episodes of spin-up and \mbox{-down}, we can infer a rough estimate of the magnetic field strength of $\approx$$1.3\times10^{12}$~G, similar to those of Her\,X--1 and 4U\,1626--67 (e.g. \citealt{bildsten97}). Such value is high with respect to the typical magnetic field inferred for radio pulsars in globular clusters.

\src\ is to our knowledge the first NS in M\,31 for which the spin period has been found. While the precise nature of the system remains unclear, it is certainly an unusual source. We propose two alternatives for \src: it could be either a peculiar LMXB pulsar (possibly within the globular cluster), similar to 4U\,1822--37 or 4U\,1626--67, or an intermediate-mass binary system akin to \mbox{Her\,X--1}, possibly observed at low inclination (given the absence of orbital  modulation of its X-ray flux). In this latter case, the analogy goes further to a similar orbital period.

\vspace{-0.5cm}

\section*{Acknowledgements} 
EXTraS is funded from the EU's Seventh Framework Programme under grant agreement n. 607452. This research is based on observations obtained with \xmm, an ESA science mission with instruments and contributions directly funded by ESA Member States and NASA. We also used data from the NASA/ESA \hst, obtained from the data archive at the STScI (operated by AURA Inc. under NASA contract NAS\,5-26555). The Digitized Sky Survey was produced at the STScI under US Government grant NAG W-2166. We are grateful to R.~Iaria for his constructive comments, and we thank R.~Gualandi for the optical spectra obtained at the 152-cm Cassini telescope at Loiano during technical nights. LS acknowledges the PRIN-INAF 2014 grant.

\vspace{-0.3cm}

\bibliographystyle{mn2e}
\bibliography{biblio}

\vspace{-0.3cm}

\section*{Supporting Information}
Additional Supporting Information may be found in the online version of this paper:\\

\noindent \textbf{Table\,\ref{obslog}.} Logbook of the \xmm\ observations used in this work.

\bsp
\label{lastpage}

\newpage
\pagestyle{empty}
\setcounter{table}{0} 
\begin{table*}
\begin{minipage}{9.3cm}
\centering \caption{Logbook of the \xmm\ observations used in this work.} \label{obslogfull}
\begin{tabular}{@{}lccccc}
\hline
Obs.\,ID & Start date & \multicolumn{2}{c}{Exposure$^{a}$}  & Off-axis$^{b}$ & Count rate$^{c}$ \\
&  & \multicolumn{2}{c}{(ks)} & (arcmin) & ($10^{-2}$~counts~s$^{-1}$) \\
\hline
0112570601 & 28-12-2000 & 13.3 & (8.8)  & 15.8 &$1.97\pm0.17$   \\
0112570101 & 06-01-2002 & 64.3 & (48.3) & 15.9 &$2.58\pm0.08$   \\
0402560901 & 26-12-2006 & 61.9 & (38.3) & 14.7 & $3.39\pm0.10$  \\
0405320701 & 31-12-2006 & 15.9 & (12.2) & 15.5 & $3.47\pm0.19$  \\
0405320801 & 16-01-2007 & 13.9 & (10.5) & 15.5 & $3.76\pm0.21$  \\
0405320901 & 05-02-2007 & 16.9 & (13.1) & 15.6 & $3.56\pm0.19$  \\
0505720201 & 29-12-2007 & 27.5 & (22.2) & 15.4 & $3.77\pm0.14$  \\
0505720301 & 08-01-2008 & 27.2 & (21.7) & 15.5 & $3.08\pm0.13$  \\
0505720401 & 18-01-2008 & 22.8 & (17.4) & 15.6 & $2.37\pm0.14$  \\
0505720501 & 27-01-2008 & 21.8 & (9.5)  & 15.5 &$3.07\pm0.21$   \\
0505720601 & 07-02-2008 & 21.9 & (17.0) & 15.6 &$3.10\pm0.16$   \\
0551690201 & 30-12-2008 & 21.9 & (15.8) & 15.4 & $3.04\pm0.15$  \\
0551690301 & 09-01-2009 & 21.9 & (16.8) & 15.5 & $3.13\pm0.15$  \\
0551690401 & 15-01-2009 & 27.1 & (3.1)  & 15.5 &$2.78\pm0.33$   \\
0551690501 & 27-01-2009 & 21.9 & (12.9) & 15.5 & $3.70\pm0.19$  \\
0551690601 & 04-02-2009 & 26.9 & (7.8)  & 15.6 &$3.03\pm0.22$   \\
0600660201 & 28-12-2009 & 18.8 & (14.4) & 15.5 &$3.50\pm0.18$   \\
0600660301 & 07-01-2010 & 17.3 & (13.2) & 15.5 & $2.58\pm0.16$  \\
0600660401 & 15-01-2010 & 17.2 & (10.3) & 15.5 &$2.53\pm0.18$   \\
0600660501 & 25-01-2010 & 19.7 & (10.2) & 15.5 & $3.67\pm0.21$  \\
0600660601 & 02-02-2010 & 17.3 & (11.5) & 15.6 & $3.01\pm0.18$  \\
0650560201 & 26-12-2010 & 26.9 & (17.0) & 15.4 & $2.10\pm0.13$  \\
0650560301 & 04-01-2011 & 33.4 & (20.3) & 15.5 &$3.34\pm0.14$   \\
0650560401 & 15-01-2011 & 24.3 & (9.2)  & 15.5 & $3.25\pm0.21$  \\
0650560501 & 25-01-2011 & 23.9 & (6.2)  & 15.6 &$2.49\pm0.24$   \\
0650560601 & 03-02-2011 & 23.9 & (14.2) & 15.6 & $2.80\pm0.16$  \\
0674210201 & 28-12-2011 & 20.9 & (16.6) & 15.4 & $1.61\pm0.12$  \\
0674210301 & 07-01-2012 & 17.3 & (11.6) & 15.5 &$1.79\pm0.14$   \\
0674210401 & 15-01-2012 & 19.9 & (14.4) & 15.5 & $1.81\pm0.13$  \\
0674210501 & 21-01-2012 & 17.3 & (13.5 & 15.6 &$1.15\pm0.12$   \\
0674210601 & 31-01-2012 & 26.0 & (15.2)& 15.6 & $1.35\pm0.11$  \\
0690600401 & 26-06-2012 &122.4 & (63.4) & 1.8  & $7.11\pm0.12$  \\
0700380501 & 28-07-2012 & 11.9 & (8.9) & 4.8  &$6.53\pm0.30$   \\
0700380601 & 08-08-2012 & 23.9 & (17.1) & 4.8  & $5.49\pm0.20$  \\
0701981201 & 08-02-2013 & 23.9 & (15.7) & 12.3 & $2.79\pm0.15$  \\
\hline
\end{tabular}
\begin{list}{}{}
\item[$^{a}$] In parentheses we give the good observing time after dead-time correction and screening for soft proton flares.
\item[$^{b}$] Radial off-axis angle of \src\ from the boresight of the pn telescope.
\item[$^{c}$] Net source count rate in the 0.3--10~keV energy band using the extraction regions described in the text; the values are not corrected for point spread function and vignetting effects.
\end{list}
\end{minipage}
\end{table*}

\end{document}